# Combined Integer and Variable Precision (CIVP) Floating Point Multiplication Architecture for FPGAs

Himanshu Thapliyal[1], Hamid R. Arabnia[2], Rajnish Bajpai[3], Kamal K. Sharma[4]

[1] NTU, Singapore; [2]The University of Georgia, USA; [3]Synopsys (India) Pvt Ltd, India; [4]G.B. Pant University, India

*Abstract*—In this paper, we propose an architecture/methodology for making FPGAs suitable for integer as well as variable precision floating point multiplication. The proposed work will of great importance in applications which requires variable precision floating point multiplication such as multi-media processing applications. In the proposed architecture/methodology, we propose the replacement of existing 18x18 bit and 25x18 bit dedicated multipliers in FPGAs with dedicated 24x24 bit and 24x9 bit multipliers, respectively. We have proved that our approach of providing the dedicated 24x24 bit and 24x9 bit multipliers in FPGAs will make them efficient for performing integer as well as single precision, double precision, and Quadruple precision floating point multiplications.

## I. INTRODUCTION

Image and digital signal processing applications require high floating point calculations throughput, and nowadays FPGAs are being used for performing these Digital Signal Processing (DSP) operations. The floating point arithmetic is of three types namely: Single precision, Double precision and Quadruple precision. The accuracy and reliability increases as we move from single precision to quadruple precision [1]. There are a number of multimedia processing applications such as graphics and geometric calculations where required degree of accuracy depends on their inputs (single precision to higher precision)[5,6]. Moreover, floating point operations are hard to implement on FPGAs as their algorithms are quite complex.

In order to combat this performance bottleneck, FPGAs vendors including Xilinx and Altera have introduced FPGAs with dedicated 18x18 bit, 25x18 bit and 9x9 bit multipliers [3,4]. These architectures can cater the need of high speed integer operations but are not that much suitable for performing floating point operations especially multiplication, which can vary from single precision to quadruple precision depending on the applications in media processing. Recently, we have addressed the issue of making FPGAs suitable for Single Precision floating point multiplication by replacing the existing 18x18 bit multiplier blocks with 24x24 bit multiplier blocks [2]. But, in literature there has not been much work for providing a unified solution of providing variable-precision floating point multiplication in FPGAs. In [7], SIMD architectures performing variable precision arithmetic blocks for FPGA based multi-media processing applications are presented. In this work, we propose a novel approach of replacing the existing 18x18 bit and 25x18 bit multipliers in FPGAs by 24x24 bit and 24x9 bit multipliers, respectively; while keeping the 9x9 bit multiplier blocks to make them viable for variable precision floating point multiplication. We have proved that dedicated 24x24 bit and 24x9 bit multipliers in FPGAs will provide efficient methodology for any precision floating point multiplication. Thus, the proposed work can also be considered as the improved solution of our work proposed in [2], which has addressed only the single precision case.

## II. PROPOSED COMBINED INTEGER AND VARIABLE PRECISION(CIVP) FLOATING POINT MULTIPLICATION ARCHITECTURE

In this section, we will demonstrate the advantage of the proposed approach of providing the dedicated 24x24 bit and 24x9 bit multipliers in FPGAs, replacing the existing 18x18 bit and 25x18 bit multipliers, while keeping the 9x9 bit multiplier blocks.

### A. Single Precision Floating Point Multiplication

In the single precision floating point algorithm, the significand of the product is determined by multiplying the two input significands(each of 23 bits) with a "1" concatenated to them. Thus, the 24x24 bit integer multiplier is the main performance bottleneck for high speed and low power operations. In FPGAs, the availability of the dedicated 18x18 bit multipliers instead of dedicated 24x24 bit multiply blocks further complicates this problem. The proposed idea of providing the dedicated 24x24 bit multipliers in FPGAs by replacing the existing 18x18 bit multipliers will provide the solution to this problem and make them efficient both for integer as well as single precision floating point multiplication operations [2].

### B. Double Precision Floating Point Multiplication

The IEEE double precision floating-point number as shown in Fig.1 consists of three fields: a 1-bit sign, *s*, an 11-bit biased exponent, *e*, and a 52-bit significand, *f*. A hidden-one bit is added to provide 53 bits of precision in normalized double precision significands.

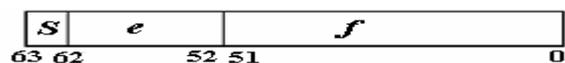

Figure 1. Double Precision Format

Now in double precision floating point multiplication, we have two mantissas A and B (each of 53 bits) for multiplication. We propose to concatenate both A and B with four bits initialized to '0'. Thus, after concatenation we have both A and B each of 57 bits, to be multiplied. Now we propose to divide the mantissas to be multiplied into three parts of 9 bit, 24 bit and 24 bit. Thus the two mantissas A and B each of 57 bits, to be multiplied, is divided into three parts; A into A1,A2 and A3 and B into B1,B2 and B3( where A1&B1 are of 9 bits each; A2 and B2 are of 24 bits; A3 and B3 are of 24 bits). The proposed approach of partitioning the mantissas A and B are also shown in Fig.2.a. It is to be noted that by our proposal in this paper, we will have availability of 24x24 bit and 24x9 bit multipliers in FPGA while 9x9 bit multipliers are already available in them. Now, the 57x57 bit mantissa multiplication can be performed as shown in Fig. 2.b (by our proposed approach of partitioning shown in Fig.2.a). As shown in Fig.2.b, we will require four 24x24 bit multipliers, four 24x9 bit multipliers and one 9x9 bit multiplier to perform this operation. Thus, by our proposed approach of providing the dedicated 24x24 and 24x9 bit multipliers along with existing 9x9 bit multipliers, will make the FPGAs efficient also for double precision floating point multiplication.

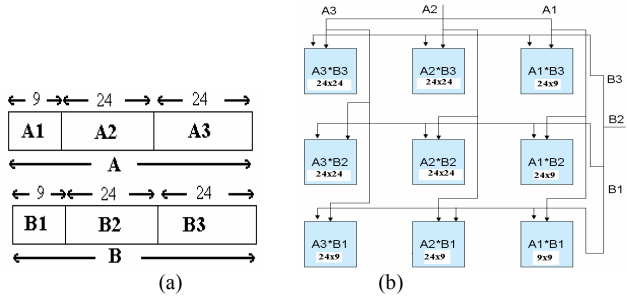

Figure 2. (a) Proposed approach of partitioning for Double precision floating point multiplication (b) 57x57 bit multiplication by the proposed approach

It is to be noted that we agree that for double precision multiplication, existing 18x18 bit multipliers in FPGAs seem to be the better choice than our proposed approach. The 54x54 bit multiplication can be achieved using nine 18x18 bit multipliers ((18 + 18 + 18 = 54). But in this paper, our aim is to address the multimedia applications where the precision requirement in floating point varies from single to higher( depending on accuracy requirement). The availability of 18x18 bit may be advisable for double precision but our proposed approach is suitable for single, double as well as quadruple precision. Thus, our proposed approach provides the unified approach of performing variable precision floating point multiplication in FPGAs.

*C. Quadruple Precision Floating Point Multiplication*

The quadruple precision number is of 128 bits and consists of a 1-bit sign, a 15-bit biased exponent, and a 112-bit significand. The quadruple precision number format is shown in Fig. 3. A hidden-one makes 113 bits of precision in normalized quadruple precision significands.

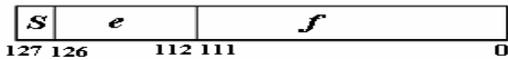

Figure 3. Quadruple Precision Format

Now in order to perform a Quadruple precision floating multiplication, a 113x113 bit integer multiplier is required. We proposed to concatenate a 1 bit initialized to '0' to both the 113 bit significands. Now we require 114x114 bit integer multiplier. We propose to divide the 114 bit mantissas A and B into A1&A2 and B1& B2 respectively (each of 57 bits), as shown in Fig.4.a. Now the 114x114 bit integer multiplication can be performed by four 57x57 bit multipliers as shown in Fig.4.b. It is to be noted that the 57x57 bit multipliers required in Fig.4.b can be implemented with our proposed approach shown in Fig.2.a and Fig. 2.b, by using four 24x24 bit multipliers, four 24x9 bit multipliers and one 9x9 bit multiplier. Our proposed scheme will work better than the existing 18x18 bit multipliers as it will require 49 18x18 bit multipliers to perform 113x113 bit multiplication (since nearest multiple of 18 greater than 113 is 126 and 126 requires 7 18 bit partitions) . If 18x18 bit multipliers are available, there will be a significant wastage of computation since we have to add 13 redundant bits initialized to '0' to 113 bits to make it 126. It shows that while partitioning the 113 bit blocks in 18 bit blocks, the last block will be of 5 bits only and 13 redundant bits will be added to make it to 18(since 113=18x6+5). Thus, out of 49 18x18 bit multipliers required to perform quadruple precision, 17 (35%) will be actually performing either 5x5 bit or 5x18 bit multiplication, while consuming the power of 18x18 bit multiplication. Our proposed approach significantly reduce this drawback as the proposed 24x24 bit, 24x9 and 9x9 multiply block will be completely utilized(will be doing the computation in their maximal capability, hence making no wastage of computation thus leading to low power). This makes our proposed approach efficient for quadruple precision multiplication in FPGAs.

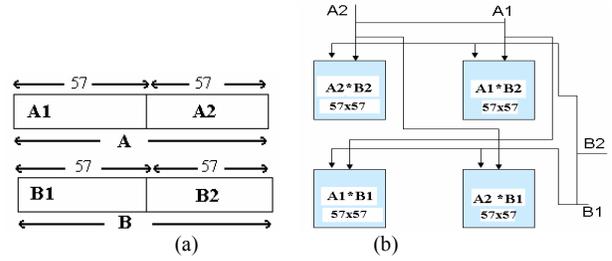

Figure 4. (a) Proposed approach of partitioning for Quadruple Floating Point Multiplication (b) Proposed 114x114 bit Multiplication

### III. ANALYSIS AND CONCLUSIONS

It is evident from the section II.A, II.B and II.C that the proposed scheme of providing dedicated 24x24 bit and 24x9 bit multipliers by replacing the existing 18x18 bit and 25x18 bit multipliers will make the FPGAs efficient for performing variable precision (Single, Double and Quadruple) floating point multiplications. The proposed scheme is also suitable for integer multiplication due to availability of the dedicated 24x24 bit multiplication blocks. The proposed scheme and its suitability for variable precision floating point multiplication are verified by coding the architectures in Verilog HDL and simulating them in ModelSim Simulator. We are also working on a novel design of 24x24 bit multiplier having the feature of reconfigurability and self reparability at run time to make the design self repairable with considerable dynamic power saving.